\documentclass[11pt,a4paper]{article}
\usepackage[utf8]{inputenc}
\usepackage{braket}
\usepackage{float}

\usepackage[sorting=none,backend=bibtex]{biblatex}
\usepackage[a4paper,width=150mm,top=25mm,bottom=25mm,bindingoffset=6mm]{geometry}
\usepackage{authblk}
\usepackage{hyperref}
\usepackage{indentfirst}
\usepackage{amsmath}
\usepackage{graphicx}
\title{\huge Improving two - qubit state teleportation affected by amplitude damping noise based on choosing appropriate quantum channel}
\author[1]{Hop Nguyen Van}
\author[1]{Anh Le Hoang}
\affil[1]{Department of Physics, Hanoi National University of Education}

 \date{}
\begin{document}
	\fontsize{12pt}{18pt}\selectfont
	\maketitle
\begin{abstract}
We consider two - qubit teleportation via quantum channel affected by amplitude damping noise. Addressing the same problem, X. Hu, Y. Gu, Q. Gong and G. Guo in Ref. \cite{xueyuan} recently showed that subjecting more qubits in quantum channel to amplitude damping can increase the fidelity of teleportation protocol. However, in this paper, by making some adjustments on quantum channel, we obtain teleportation fidelity which is even higher than one in the procedure of X. Hu \textit{et al}. Moreover, our strategy is simpler than quantum distillation and compared to using weak measurement, it is deterministic. Furthermore, explicit analysis of fidelity is provided, we show that in general, choosing appropriate quantum channel enhances the ability of teleportation better and negates the fact that \textit{more amplitude damping noise more quality}. 

\end{abstract}

\section{Introduction}
Quantum teleportation, firstly proposed by Bennett \textit{et al} \cite{teleportation}, is a protocol that transports an unknown quantum state from a sender (Alice) to a receiver (Bob) without transferring the physical carrier of the state. To complete this task, Alice and Bob at the beginning need to share a maximal entangled state which serves as the quantum channel. Then, Alice performs a two – qubit measurement on her qubits and sends the results to Bob through classical communication. Based on Alice’s outcomes, Bob applies appropriate unitary operations to obtain the desired state. Later, teleportation was developed to a two - qubit one. In 2002, Lee \textit{et al} \cite{lee2002} showed that it was possible to teleport a two - qubit state $\left| \Phi  \right\rangle  = a\left| {00} \right\rangle  + b\left| {01} \right\rangle  + c\left| {10} \right\rangle  + d\left| {11} \right\rangle $ from Alice to Bob using a four entangled state. After that, G. Rigolin \cite{rigolin2005} explicitly constructed the protocol and created 16 Generalized Bell states to implement the two - qubit teleportation.\\
\indent In general, to achieve faithful teleportation, the quantum channel connecting Alice and Bob is considered to be maximally entangled. However, due to coupling with environment, an entangled state may lose its coherence and become a mixed one, whose consequence is to reduce the efficiency of the protocol. It is natural to think that in order to obtain secure teleportation, one must "fix" the noisy quantum channel. In this perspective, the first strategy, quantum distillation, is introduced in Ref. \cite{distillation}. In quantum distillation, entangled pairs are purified - converted to almost perfectly entangled states from mixed states. Such purification can be accomplished by performing local unitary operations and measurements on the shared entangled pairs. Another solution is using weak measurement to improve the fidelity of teleportation via noisy channels \cite{weakmeasurement}. However, these techniques have disadvantages in which the former requires heavy quantum technology and the latter is non trace - preserving (achieves success lower than 100$\% $).\\
\indent Nevertheless, in contrast with changing the noisy quantum channel to match with original teleportation (quantum distillation and weak measurement), there is another idea suggesting that the standard teleportation should be altered to be compatible with the affected quantum channel \cite{albeverio,ghosh,fighting}. In comparison with the preceding methods, this strategy is trace - preserving and does not destroy the entangled pairs. From this point of view, Bandyopadhyay in Ref. \cite{bandyopadhyay} showed that a higher single - qubit teleportation fidelity can be achieved even when more qubits of quantum channel are affected by amplitude damping noise. A similar effect was also pointed out for two - qubit teleportation in which adding more amplitude damping noise to Bell pair employed as quantum channel can enhance the fidelity \cite{xueyuan}. Such results imply that \textit{more amplitude damping noise more quality} and they are counter - intuitive in the sense that more amplitude damping noise in principle, reduces entanglement of quantum channel and makes teleportation protocol worst (\textit{more amplitude damping noise worst quality}). Yet, in this paper we provide a case which goes against the preceding results, we also show that the fidelity of two - qubit teleportation in case of amplitude damping noise can be enhanced more by making legitimate adjustments in the quantum channel.\\
\indent The remaining of this paper is organized as follows. In section 2,  we introduce quantum teleportation of two - qubit state affected by amplitude damping noise while section 3 is devoted to mathematical results and discussion. The final conclusion is given in section 4.

\section{Quantum teleportation of two – qubit state affected by amplitude damping noise}
The preparation of two - qubit teleportation could be described as follows. Suppose that the arbitrary two - qubit state Alice wants to teleport to Bob has the following form ${\left| \psi  \right\rangle _{{X_1}{X_2}}} = {\left( {a\left| {00} \right\rangle  + b{e^{i{\varphi _1}}}\left| {01} \right\rangle  + c{e^{i{\varphi _2}}}\left| {10} \right\rangle  + d{e^{i{\varphi _3}}}\left| {11} \right\rangle } \right)_{{X_1}{X_2}}}$, here the real parameters $a$, $b$, $c$, $d$, ${\varphi _1}$, ${\varphi _2}$ and ${\varphi _3}$ satisfy the normalization condition $a^2+b^2+c^2+d^2 = 1$. Hereafter, this state is called the input state. The quantum channel shared between Alice and Bob is ${\left| Q \right\rangle _{{A_1}{B_1}{A_2}{B_2}}} = \mathop  \otimes \limits_{i = 1}^2 {\left( {\cos {\theta _i}\left| {00} \right\rangle  + \sin {\theta _i}\left| {11} \right\rangle } \right)_{{A_i}{B_i}}}$ which is also known as a tensor product of two Bell - like states. Its density matrix could be written as
\begin{equation}
{\rho _{ch}} = {\left| Q \right\rangle _{{A_1}{B_1}{A_2}{B_2}}}\left\langle Q \right| = \mathop  \otimes \limits_{i = 1}^2 \left( {\begin{array}{*{20}{c}}
{{{\cos }^2}{\theta _i}}&0&0&{\sin {\theta _i}\cos {\theta _i}}\\
0&0&0&0\\
0&0&0&0\\
{\sin {\theta _i}\cos {\theta _i}}&0&0&{{{\sin }^2}{\theta _i}}
\end{array}} \right).
\end{equation}
\indent Here subindex \textit{ch} means "channel", the qubits $A_{1}, A_{2}$ ($B_{1},B_{2}$) are held by Alice (Bob). Note that when ${\theta _1} = {\theta _2} = {\pi  \mathord{\left/
 {\vphantom {\pi  4}} \right.
 \kern-\nulldelimiterspace} 4}$, the quantum channel transforms into a tensor product of two Bell states, $\left| {{\phi ^ + }} \right\rangle  \otimes \left| {{\phi ^ + }} \right\rangle $. Here free parameters $\theta_{1}$ and $\theta_{2}$ are introduced to optimize the teleportation performance. The first step of the protocol is that Alice performs two Bell measurements on the pairs of ($X_{1},A_{1}$) and ($X_{2},A_{2}$). After the measurements, she publicly announces the results. In the last step, according to the measurement outcomes of Alice, Bob applies corresponding local unitary operations to his qubits to recover the desired state.\\
\indent Here we consider the scenario that after the preparation of quantum channel, each pair of qubits ($A_{1}, A_{2}$ and $B_{1},B_{2}$) are sent to Alice and Bob via two independent amplitude damping channels. The physical meaning of amplitude damping noise relates to the interaction of a two - level atom in the electromagnetic field. Particularly, amplitude damping noise describes the decay of an excited state of a two - level atom by emission of a photon or losing a quantum energy to the environment in presence of an electromagnetic field. The effect of amplitude damping on a qubit can be conveniently described by Kraus operators \cite{nielsen}
\begin{equation}
{E_1} = \left( {\begin{array}{*{20}{c}}
1&0\\
0&{\sqrt {1 - p} }
\end{array}} \right),\quad {E_2} = \left( {\begin{array}{*{20}{c}}
0&{\sqrt p }\\
0&0
\end{array}} \right),
\end{equation}
with $p$ is the probability of emitting photon, or specifically the noise strength $\left( {0 \le p \le 1} \right)$. After being subjected to amplitude damping channels, the initial state ${\rho _{ch}}$ becomes a mixed one which could be represented in terms of operator - sum representation
\begin{equation}
{\rho '_{ch}} = \sum\limits_{i,j,m,n} {{E_{ijmn}}\left( {{p_A},{p_B}} \right){\rho _{ch}}E_{ijmn}^\dag \left( {{p_A},{p_B}} \right)} ,
\end{equation}
where ${E_{ijmn}}\left( {{p_A},{p_B}} \right)$ are the Kraus operators acting on whole quantum channel and $p_{A}$ $(p_{B})$ is the strength of amplitude damping noise acting on qubits of Alice (Bob).\\
\indent The efficiency of the protocol can be evaluated through teleportation fidelity. The higher teleportation fidelity is, the more effective  the protocol is. Since the initial state is pure, the fidelity could be written as
\begin{equation}
f = {}_{{X_1}{X_2}}\left\langle \psi  \right|{\rho _B}{\left| \psi  \right\rangle _{{X_1}{X_2}}},
\end{equation}
with $\rho_{B}$ being the state which Bob obtains in the last step of teleportation protocol. To obtain the average fidelity of all possible input states, we set $a = \cos {\eta _3}$, $b = \sin {\eta _3}\cos {\eta _2}$, $c = \sin {\eta _3}\sin {\eta _2}\cos {\eta _1}$ and $d = \sin {\eta _3}\sin {\eta _2}\sin {\eta _1}$. According to Ref. \cite{averagefidelity}, the formula of average fidelity takes the following form
\begin{equation}
\left\langle F \right\rangle  = \frac{{3!}}{{{\pi ^3}}}\prod\limits_{k = 1}^3 {\int\limits_0^{{\pi  \mathord{\left/
 {\vphantom {\pi  2}} \right.
 \kern-\nulldelimiterspace} 2}} {\cos {\eta _k}{{\left( {\sin {\eta _k}} \right)}^{2k - 1}}d{\eta _k}} \prod\limits_{k = 1}^3 {\int\limits_0^{2\pi } {d{\varphi _k}f,} } } 
\end{equation}
where ${\eta _k} \in \left[ {0,{\pi  \mathord{\left/
 {\vphantom {\pi  2}} \right.
 \kern-\nulldelimiterspace} 2}} \right]$ and ${\varphi _k} \in \left[ {0,2\pi } \right)$.

\section{Results and discussion}
The analytical expression of average fidelity we obtained reads
\begin{equation}
\left\langle F \right\rangle  = \frac{1}{5} + \frac{1}{5}\prod\limits_{k = 1}^2 {\left[ {1 + \left( {2{p_A}{p_B} - {p_A} - {p_B}} \right){{\sin }^2}{\theta _k} + \sqrt {\left( {1 - {p_A}} \right)\left( {1 - {p_B}} \right)} \sin 2{\theta _k}} \right]} .
\label{tongquat}
\end{equation}
\indent First of all, by setting ${\theta _1} = {\theta _2} = {\pi  \mathord{\left/
 {\vphantom {\pi  4}} \right.
 \kern-\nulldelimiterspace} 4}$, we recover the results of using Bell pairs as quantum channel in Ref. \cite{xueyuan}, in which X. Hu \textit{et al} claimed that adding more noise (to Alice's qubits) can enhance ability of two - qubit state teleportation. When qubits of both Alice and Bob are subjected to noise, from Eq. (\ref{tongquat}), we could obtain the corresponding fidelity
\begin{equation}
\left\langle {F_{AB}^{Bell}} \right\rangle  = \frac{1}{5} + \frac{1}{5}{\left( {1 + {p_A}{p_B} - \frac{{{p_A}}}{2} - \frac{{{p_B}}}{2} + \sqrt {\left( {1 - {p_A}} \right)\left( {1 - {p_B}} \right)} } \right)^2},
\end{equation}
here the subindex means the qubits are acted by noise ($A$ means $A_{1}$, $A_{2}$ and $B$ means $B_{1}$, $B_{2}$) and the superindex means the type of quantum channel. When only Bob’s qubits are subjected to noise, we set $p_{A}=0$ and obtain
\begin{equation}
\left\langle {F_B^{Bell}} \right\rangle  = \frac{1}{{20}}\left( {6 + 4\sqrt {1 - {p_B}}  - {p_B}} \right)\left( {2 - {p_B}} \right).
\end{equation}
\begin{figure}[H]
	\centering
	\includegraphics[scale=0.5]{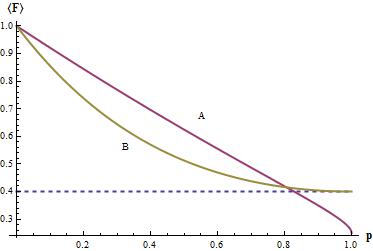}
	\caption{Average fidelity as a function of noise strength. Curves A and B are, respectively, the evolution of $\left\langle {F_B^{Bell}} \right\rangle $ and $\left\langle {F_{AB}^{Bell}} \right\rangle $ . The horizontal dashed line at 2/5 represents the classical limit.}
	\label{figure1}
\end{figure}
\indent The dependence of $\left\langle {F_B^{Bell}} \right\rangle $ ($\left\langle {F_{AB}^{Bell}} \right\rangle $) on noise strength $p$ is plotted as curve A (curve B) in Fig. \ref{figure1} (here $p_A=p_B=p$). For high values of $p$ (when $p$ is approximately greater than 0.8), adding more noise to quantum channel is really effective, since $\left\langle {F_B^{Bell}} \right\rangle $ is lower than the classical limit ${2 \mathord{\left/
 {\vphantom {2 5}} \right.
 \kern-\nulldelimiterspace} 5}$ while $\left\langle {F_{AB}^{Bell}} \right\rangle $  always exceeds ${2 \mathord{\left/
 {\vphantom {2 5}} \right.
 \kern-\nulldelimiterspace} 5}$. However, for low values of $p$, curve B is below curve A so the scenario that only Bob’s qubits are subjected to noise now achieves better fidelity, adding more noise does not work anymore. To sum up, adding more noise is only valid for high values of noise strength.\\
\indent At this point, we found that by setting appropriate values of $\theta_k$ ($k = 1,2$), we could achieve optimal average fidelity better than both ${2 \mathord{\left/
 {\vphantom {2 5}} \right.
 \kern-\nulldelimiterspace} 5}$ and $\left\langle {F_B^{Bell}} \right\rangle $. The optimal values $\theta _k^{opt}$ that make average fidelity $\left\langle F \right\rangle $ maximal can be determined from the equations ${\left. {{{\partial \left\langle F \right\rangle } \mathord{\left/
 {\vphantom {{\partial \left\langle F \right\rangle } {\partial {\theta _k}}}} \right.
 \kern-\nulldelimiterspace} {\partial {\theta _k}}}} \right|_{{\theta _k} = \theta _k^{opt}}} = 0$ and the conditions ${\left. {{{{\partial ^2}\left\langle F \right\rangle } \mathord{\left/
 {\vphantom {{{\partial ^2}\left\langle F \right\rangle } {\partial \theta _k^2}}} \right.
 \kern-\nulldelimiterspace} {\partial \theta _k^2}}} \right|_{{\theta _k} = \theta _k^{opt}}} < 0$. Solving these equations with the conditions we have
\begin{equation}
\theta _k^{opt} = \frac{1}{2}\arctan \frac{{2\sqrt {\left( {1 - {p_A}} \right)\left( {1 - {p_B}} \right)} }}{{{p_A} + {p_B} - 2{p_A}{p_B}}}.
\label{theta}
\end{equation}
\indent With so chosen values of $\theta _k^{opt}$ we derive the optimal average fidelity ${\left\langle {F_{AB}^{Bell - like}} \right\rangle _{opt}}$. The analytical expression reads
\begin{equation}
\begin{array}{l}
{\left\langle {F_{AB}^{Bell - like}} \right\rangle _{opt}} = \frac{1}{5}\\
\quad  + \frac{1}{{20}}{\left( {2 - {p_A} - {p_B} + 2{p_A}{p_B} + \sqrt {4\left( {1 - {p_A}} \right)\left( {1 - {p_B}} \right) + {{\left( {{p_A} + {p_B} - 2{p_A}{p_B}} \right)}^2}} } \right)^2}.
\end{array}
\end{equation}
Set $p_A=0$ we obtain
\begin{equation}
{\left\langle {F_B^{Bell - like}} \right\rangle _{opt}} = \frac{1}{5} + \frac{1}{5}{\left( {2 - {p_B}} \right)^2}.
\end{equation}
\begin{figure}
	\centering
	\includegraphics[scale=0.7]{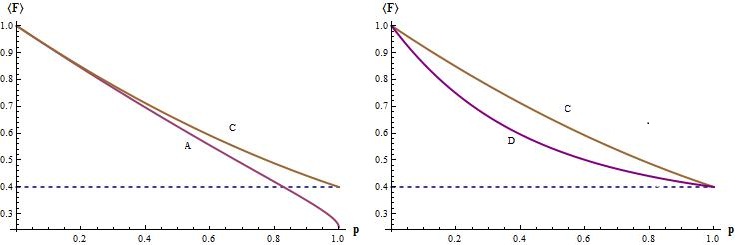}
	\caption{Average fidelity as a function of noise strength. Curve A, C and D are, respectively, the evolution of $\left\langle {F_B^{Bell}} \right\rangle $, ${\left\langle {F_B^{Bell - like}} \right\rangle _{opt}}$ and $\left\langle {F_{AB}^{Bell - like}} \right\rangle _{opt} $. The horizontal dashed line at 2/5 represents the classical limit.}
	\label{figure2}
\end{figure}
\indent The dependence of ${\left\langle {F_B^{Bell - like}} \right\rangle _{opt}}$ on noise strength $p$ ($p_B=p$) is plotted in Fig. \ref{figure2}, which is illustrated as curve C. It can be observed that for all possible values of $p$, both curve A and classical limit line are always below curve C. So the first remarkable result we obtained is to point out that, choosing appropriate parameters $\theta_1$ and $\theta_2$, or equivalently appropriate quantum channel achieves better fidelity than adding more noise.\\
\indent Moreover, by making some analyses on ${\left\langle {F_{AB}^{Bell - like}} \right\rangle _{opt}}$ plotted as curve D in Fig. \ref{figure2}, we found a case against the controversial idea of \textit{more amplitude damping noise more quality} in Ref. \cite{xueyuan}. Although the optimal values of parameters $\theta_k$ are chosen in Eq. (\ref{theta}), for all values of noise strength $p$, curve D lies below curve C. It suggests that in this case, adding more noise always decreases the teleportation fidelity. Hence, our result satisfies the usual idea of \textit{more amplitude damping noise worst quality}.

\section{Conclusion}
To conclude, we have investigated a strategy to obtain higher teleportation fidelity than adding more noise: choosing appropriate quantum channel. In comparison with previous strategies including quantum distillation, weak measurement and adding more noise, our strategy is simpler, deterministic (achieve 100$\%$ of success) and effective for all values of noise strength. Furthermore, we have also provided a case that more amplitude damping noise decreases the ability of two - qubit state teleportation, which is not in agreement with the idea of \textit{more amplitude damping noise more quality} in Ref. \cite{xueyuan}. We hope that our results could be useful in practical teleportation.

\printbibliography

\end{document}